\documentclass{article}
\usepackage{spconf,amsmath,graphicx,amssymb,multirow,tabularx,array,booktabs,fancyhdr,hyperref,adjustbox}

\title{Enhancing Quantised End-to-End ASR Models via Personalisation}
\name{Qiuming Zhao\textsuperscript{1}, Guangzhi Sun\textsuperscript{2}, Chao Zhang\textsuperscript{1}, Mingxing Xu\textsuperscript{1}, Thomas Fang Zheng\textsuperscript{1}$^{\ast}$\thanks{$\ast$Correspondence}}
\address{\textsuperscript{1}Tsinghua University, China; \textsuperscript{2}University of Cambridge, United Kingdom\\
\texttt{\small{zqm23@mails.tsinghua.edu.cn; gs534@cam.ac.uk; \{cz277,xumx,fzheng\}@tsinghua.edu.cn}}}
\fancypagestyle{specialfooter}{%
  \fancyhf{} 
  \fancyfoot[LE,LO]{%
    \begin{minipage}{\textwidth}
      \rule[1ex]{0.2\textwidth}{0.75pt} \\ 
      \small\href{https://github.com/qmgzhao/PQM.git}{\textsuperscript{1}Data partition details: https://github.com/qmgzhao/PQM.git} 
    \end{minipage}
  }
}

\begin{document}
\ninept
\maketitle
\begin{abstract}
Recent end-to-end automatic speech recognition (ASR) models have become increasingly larger, making them particularly challenging to be deployed on resource-constrained devices. Model quantisation is an effective solution that sometimes causes the word error rate (WER) to increase. In this paper, a novel strategy of personalisation for a quantised model (PQM) is proposed, which combines speaker adaptive training (SAT) with model quantisation to improve the performance of heavily compressed models. Specifically, PQM uses a 4-bit NormalFloat Quantisation (NF4) approach for model quantisation and low-rank adaptation (LoRA) for SAT. Experiments have been performed on the LibriSpeech and the TED-LIUM 3 corpora. Remarkably, with a 7x reduction in model size and 1\% additional speaker-specific parameters, 15.1\% and 23.3\% relative WER reductions were achieved on quantised Whisper and Conformer-based attention-based encoder-decoder ASR models respectively, comparing to the original full precision models.
\end{abstract}
\section{Introduction}
\label{sec:intro}

End-to-end neural network models have achieved state-of-the-art results in various Automatic Speech Recognition (ASR) tasks \cite{radford2023robust, gulati2020conformer, zhang2020transformer, kriman2020quartznet}. However, these advancements in accuracy usually come at the cost of increasing model size, which not only substantially increases operational costs on the server but also presents significant challenges in deploying them on resource-constrained edge devices. More recently, universal large speech models, such as OpenAI Whisper \cite{radford2023robust} have become increasingly popular. These models adopt a very large amount of model parameters, making the deployment even more challenging and demanding.

Model quantisation has been widely adopted as an effective way to reduce model sizes, and has been widely studied and applied in both academia and industry \cite{han2015deep, qian2019binary, leng2018extremely}. Model quantisation replaces floating-point weights with low-precision values to considerably reduce the model size and inference time without altering the model architecture. However, model quantisation often results in a degradation in model performance due to the loss of precision. Some studies employ quantisation-aware training (QAT) schemes \cite{nguyen2020quantization, ding20224, zhen2022sub} that consider the effects of quantisation during training, while others utilize more sophisticated quantisation methods \cite{xu2022towards, hernandez2023sharing, yao2020int8}. Although these approaches have mitigated the performance degradation issue directly from the generic data perspective, the fact that the edge devices to deploy quantised models are often personalised is under-explored. 
For these devices, such as personalised voice assistants or smart door locks, improving performance for the target speaker is the critical objective rather than the generic performance. 
Consequently, this paper investigates the use of personalisation to compensate for the degradation due to quantisation.

This paper proposes a novel strategy of personalisation for a quantised model (PQM) which performs speaker adaptive training on a quantised end-to-end ASR model. The PQM strategy adopts the block-wise NormalFloat4 (NF4) quantisation \cite{dettmers2023qlora} for model compression, which incurs a smaller performance loss compared to conventional uniform quantisation. The speaker adaptive training is performed using the Low-Rank Adaptation (LoRA) \cite{hu2021lora} approach. As the adaptation data for a particular speaker is often very limited, PQM includes a LoRA pretraining stage before the speaker adaptive training using existing similar data to achieve faster and better convergence. Moreover, to further mitigate the data scarcity issue, semi-supervised training is also explored in PQM where the quantised model can be trained with labels generated by the model itself.


The PQM strategy was implemented for the Conformer attention-based encoder-decoder (AED) model and the Whisper model as two examples of end-to-end ASR models in this paper. Experiments performed on the LibriSpeech and TED-LIUM 3 datasets demonstrated that, with nearly a 7-fold compression of the model, the PQM strategy achieves a relative WER reduction of 15.1\% and 23.3\% on quantised Whisper and Conformer AED models respectively, compared to the original full-precision models. The main contribution of this paper can be summarised as follows.
\begin{itemize}
\setlength\itemsep{0.1em}
    \item The PQM strategy, to the best of our knowledge, is the first work that investigates personalisation to compensate for quantisation degradation on edge devices.
    \item A LoRA pretraining-based initialisation and a semi-supervised training approach are proposed for speaker adaptive training.
    \item PQM has been validated on both Conformer AED and Whisper models across two datasets, with large improvements achieved compared to fully fine-tuned baselines.
\end{itemize}

The rest of this paper is organised as follows: Sec. \ref{sec:related work} summarises related work. Sec. \ref{sec:methodology} explains the three stages and some details of the PQM strategy. Sec. \ref{sec:experiments} describes the experimental setup and results are provided in Sec. \ref{sec:results}. The paper concludes in Sec. \ref{sec:conclusions}.

\begin{figure*}[t]
  \centering
  \includegraphics[width=0.9\linewidth]{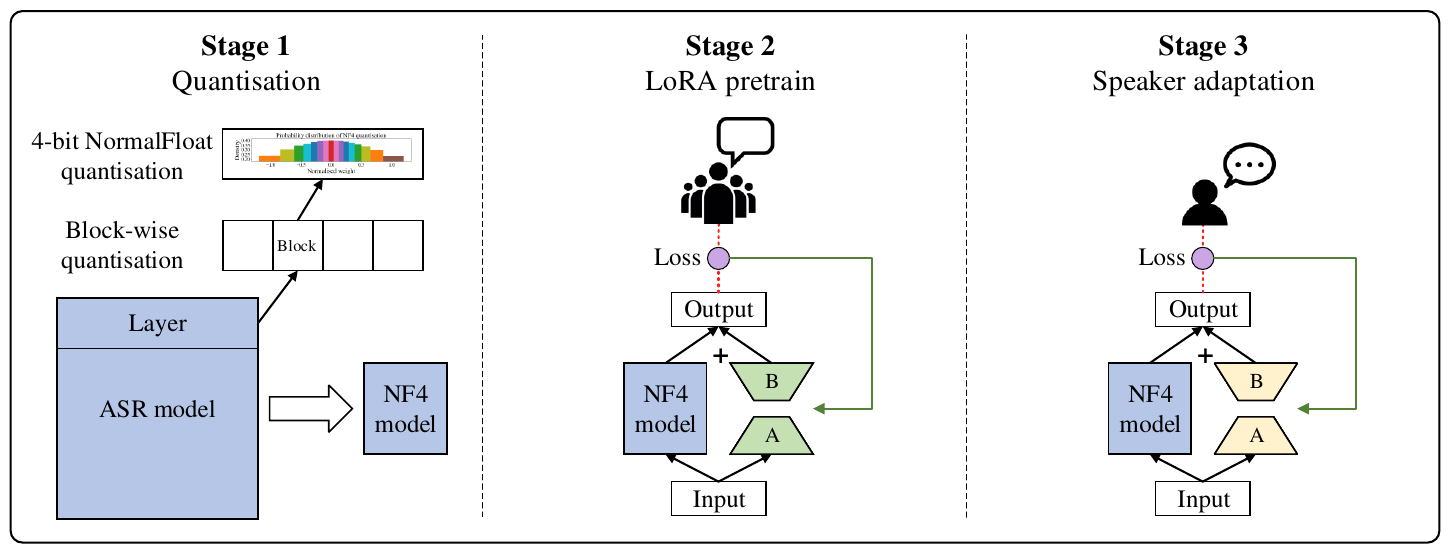}
  \vspace{-0.5cm}
  \caption{Overview of the PQM strategy.}
  \label{fig:Strategy}
\end{figure*}

\section{Related Work}
\label{sec:related work}

\subsection{Quantisation}
\label{ssec:related work quantisation}

Quantisation is the process of discretising an input with higher information content to one with lower information content. Uniform quantisation is a commonly used quantisation method where the value range is uniformly divided into multiple intervals. This approach is widely supported in mobile inference frameworks such as TFLite \cite{tensorflow_lite}, MNN \cite{jiang2020mnn}, and NCNN \cite{ncnn}. Recently, Quantile Quantisation was proposed based on the \textit{Lossy Minimum Entropy Encoding} \cite{dettmers20218}, and QLoRA was introduced \cite{dettmers2023qlora} which employs NF4 quantisation on the weights of pretrained neural networks to achieve near-lossless quantisation. 
Unlike uniform quantisation, Quantile Quantisation determines the quantisation steps and levels based on the data distribution. In data-dense regions, more quantisation levels are allocated, ensuring an equal number of quantised values in each quantisation bin. This allows for a more refined representation of the data, as well as better handling of outliers and adaptation to non-uniform distributions.

\subsection{Speaker Adaptation}
\label{ssec:speaker adaptation}

The objective of speaker adaptation is to minimize the mismatch between speakers in training and testing conditions. Current neural network-based methods for speaker adaptation can be broadly categorized into two types: Embedding-based and Model-based.


Speaker embeddings map speakers into a continuous space using techniques like i-vectors \cite{saon2013speaker, sari2020unsupervised, rohdin2018end} or neural network bottlenecks \cite{yue2020autoencoder, cardinal2015speaker, doddipatla2014speaker}.
Model-based adaptation methods include three primary methods: Structured Transforms, Speaker Adaptive Training (SAT), and Regularization.
Structured Transforms, such as the Learning Hidden Unit Contributions (LHUC) scheme \cite{swietojanski2016learning} and parameterised activation functions \cite{zhang2016dnn}, modify the model structure or its activations.
SAT methods adjust model parameters to individual speakers using approaches like SAT-embedding \cite{miao2015speaker} and SAT-LHUC \cite{swietojanski2016sat}.
Regularization techniques, such as L2 loss or KL divergence \cite{liao2013speaker, yu2013kl}, work to prevent overfitting to specific speakers.

\section{Methodology}
\label{sec:methodology}

\subsection{Strategy Overview}
\label{ssec:strategy overview}

The PQM strategy is illustrated in Fig. \ref{fig:Strategy}, which is divided into three stages. In stage 1, we apply block-wise NF4 quantisation to the model's primary weight parameters. In stage 2, we pretrain the randomly initialised LoRA using data from a large number of speakers, providing a more robust starting point for subsequent speaker adaptation. In stage 3, we perform speaker adaptive training on speaker-specific data, during which the entire model is frozen, and only the LoRA parameters corresponding to each speaker are updated.

Stage 2 of PQM is particularly beneficial for the application scenario where reasonable-sized training data of the target domain is available, e.g. assuming there is some in-house data, with a very limited amount of target speaker data.

\subsection{k-bit NormalFloat Quantisation}
\label{ssec:quantisation}

The block-wise NF4 quantisation is adopted in this paper, which is applied to the weight matrices that are the primary parameters of the model. While standard floating point quantisation applies the same set of quantisation bins to all weight matrices, the dynamic range of parameter values is not taken into account, resulting in heavily unbalanced quantisation bins. NF4, on the contrary, ensures each bin has an equal number of values by estimating the quantile of the input matrices using the empirical cumulative normal distribution. This leveraged the fact that the parameters of a weight matrix, in general, follow a normal distribution \cite{dettmers2023qlora}. 

\begin{figure}[t]
  \centering
  \includegraphics[width=\linewidth]{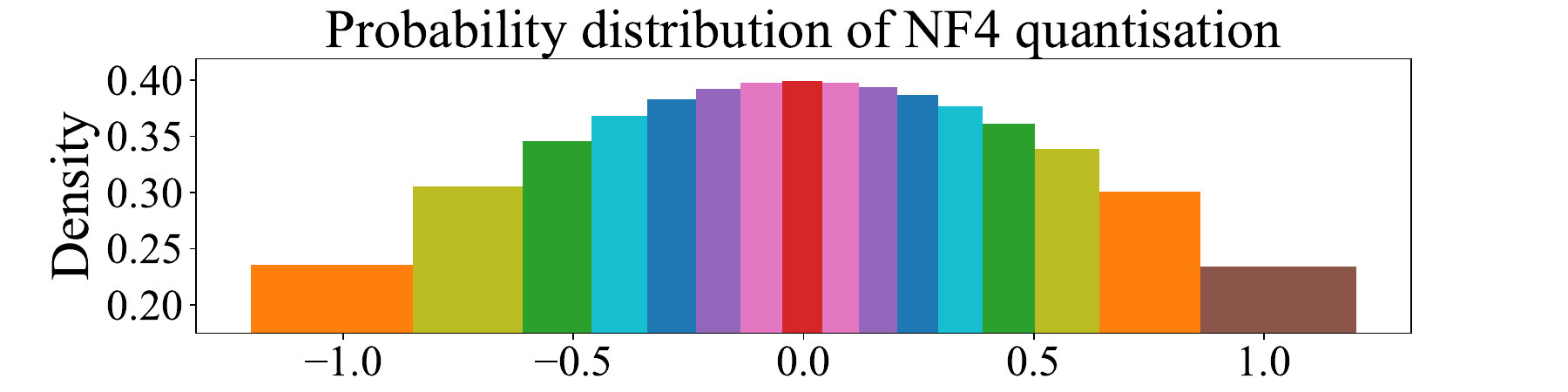}
  \vspace{-0.3cm}
  \caption{\textit{Illustration of the construction of quantiles for NF4 quantisation. It comprises 16 quantisation bins, where the midpoint of each bin represents the quantisation level.}
  \vspace{-0.3cm}
}
  \label{fig:NF4}
\end{figure}
Specifically, for a $k$-bit quantisation, $2^k+1$ quantiles (i.e. $2^k$ quantisation bins) of a theoretical $\mathcal{N}(0, 1)$ distribution are first estimated which equally divides the area under the distribution curve. Then, these quantisation bins are normalised to be within the range $[-1, 1]$, as illustrated in Fig. \ref{fig:NF4}. Finally, the parameters of a weight matrix are normalised into the range $[-1, 1]$ to find their corresponding quantiles by dividing the maximum absolute value of those parameters. In this way, a similar number of quantised values are obtained for each bin, allowing for a more refined representation of the model parameters. To ensure zero is exactly quantised to zero which is important for padding, asymmetric quantiles are used that ensure the mid bin has the quantisation level of zero (see Fig. \ref{fig:NF4}).

To reduce the influence of extreme values in weight matrices (i.e. outliers) on the maximum absolute value normalisation, block-wise quantisation is applied which divides the weight matrices into small blocks and quantises each block with separate normalisation factors.
In this way, outliers in the input tensor are confined to individual blocks, reducing their overall impact on quantisation. As a result, block-wise quantisation allows for individual normalisation factors for each block, resulting in a more fine-grained overall quantisation.




\subsection{LoRA for Speaker Adaptation}
\label{ssec:LoRA for speaker adaptation}

Compared to full fine-tuning, LoRA adjusts only the low-rank subspace parameters of the model, thereby achieving higher computational efficiency and lower costs for computation and storage. In scenarios with limited speaker data, full fine-tuning methods may be prone to model overfitting, whereas LoRA can alleviate this issue.

\begin{figure}[t]
  \centering
  \includegraphics[scale=0.7]{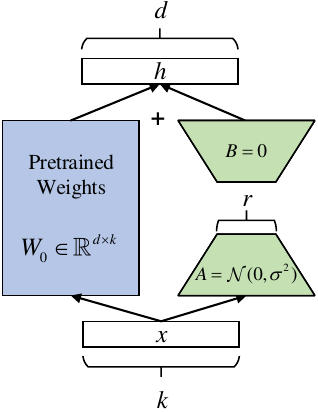}
  \vspace{-0.3cm}
  \caption{\textit{LoRA. Initially, the pretrained weight parameters \( W_0 \) are frozen. For \( A \), random Gaussian initialisation is employed, whereas \( B \) is initialised with zeros.}}
  \label{fig:LoRA}
  \vspace{-0.3cm}
\end{figure}
For the pretrained ASR model with weight matrix \( W_0 \in \mathbb{R}^{d \times k} \), its update is expressed through the following equation, where \( B \in \mathbb{R}^{d \times r} \) and \( A \in \mathbb{R}^{r \times k} \), with the rank \( r \ll \min(d, k) \).
\begin{equation}
W_0 + \Delta W = W_0 + B A
\end{equation}
In speaker adaptive training, only the LoRA parameters corresponding to each speaker are updated. In this way, effective adaptation to different speakers can be achieved by updating a minimal set of parameters. Moreover, in cases where the base model is quantised, full-precision LoRA serves to some extent to restore full-precision performance to the base model.

Although the target speaker data is always limited, in reality, the target domain data of other speakers is usually available. Therefore, PQM leverages those data to find a better initialisation point for LoRA weights before performing speaker adaptation, referred to as LoRA pretraining.
This allows the LoRA parameters to adapt to the target-domain ASR task, providing a more robust starting point for subsequent speaker adaptation.
When using LoRA, each speaker corresponds to one set of LoRA parameters. The number of parameters is given by \( |\Theta| = 2 \times L_{\text{LoRA}} \times d_{\text{model}} \times r \), where \( L_{\text{LoRA}} \) is the number of weight matrices to apply LoRA, \( d_{\text{model}} \) is the attention layer dimension, and \( r \) is the rank.

\section{Experimental Setup}
\label{sec:experiments}

\subsection{Data}
\label{ssec:dataset}


{\bf LibriSpeech} is an English audiobook dataset. We selected 5 male speakers and 5 female speakers with the largest number of utterances from train-clean-360 as speaker adaptation data. Each speaker contributes approximately 150 utterances, resulting in a total speech duration of roughly 25 minutes. For LoRA pre-training, the train-clean-100 set was used which does not have any speaker overlap with the selected speakers.

{\bf TED-LIUM 3} (TL3) is a TED talks dataset. We selected 16 speakers from the test set as speaker adaptation data. On average, each speaker has 161 utterances (14 minutes).

Speaker adaptation data for LibriSpeech and TL3 was divided randomly, where 2/5 was divided into the train set, 1/5 was divided into the dev set, and 2/5 was divided into the test set. On average, each speaker has 6-10 minutes of training data, while the dev and test data remains constant across all experiments. We denote the partitioned test sets as \textit{LibriSpeech-SA} and \textit{TL3-SA} respectively in the results. Data partition details are provided\textsuperscript{1}.
\thispagestyle{specialfooter} 

\subsection{Model and training specifications}
\label{ssec:baseline}

In order to verify the effectiveness of PQM, we use Whisper and Conformer AED models as two widely used models as examples.

{\bf Whisper} is a transformer-based AED model released by OpenAI trained on 680k hours of audio. The base.en model with a full model size of 278MB was used. The encoder has 6 Transformer blocks with 2048 hidden dimensions, and the output size is 512. The decoder has 6 Transformer blocks with 2048 hidden dimensions. The Transformer-related weight matrices are all 512 by 512 dimensional. Feature processing and model training followed \cite{radford2023robust,whisperbiasing}.

{\bf Conformer AED} is a hybrid CTC/attention-based encoder-decoder model, whose FP32 model size is about 131MB. The training follows ESPnet \cite{espnet} with 0.3 CTC weight and 80-dim FBank features. The Conformer encoder has 12 blocks with 1024 hidden dimensions. The decoder uses a 6-block transformer architecture with 2048-dim linear units. The Transformer-related weight matrices are all 256 by 256.

The baseline system is the quantised system without personalisation. When training LoRA from scratch (rank=4), the LoRA parameter sizes of Whisper and Conformer are 1.2MB and 0.8MB. When using pretrained LoRA (rank=1), the LoRA parameter sizes of Whisper and Conformer are 0.3MB and 0.2MB. Identical hyper-parameter settings were used for all speakers. Models are evaluated using WER averaged across all utterances from the test set speakers.

\section{Evaluation Results and Analysis}
\label{sec:results}


\newcommand{\tabincell}[2]{\begin{tabular}{@{}#1@{}}#2\end{tabular}}

\begin{table}[t]
\caption{\textit{WER on the LibriSpeech-SA using quantised Whisper and Conformer for the linear, convolution, and embedding layers.}}
\vspace{0.2cm}
    \begin{adjustbox}{width=\columnwidth,center}
    \setlength{\tabcolsep}{2.5pt}
    \centering
    \begin{tabular}{lcccccc}  
        \toprule
        \multirow{2}{*}[-0.25em]{\parbox{1.2cm}{\centering \textbf{System}}} & \multicolumn{3}{c}{\textbf{Quantise}} & \multirow{2}{*}[-0.25em]{\textbf{WER(\%)}} & \multirow{2}{*}[-0.25em]{\textbf{\tabincell{c}{Model Size\\(MB)}}} & \multirow{2}{*}[-0.25em]{\textbf{\tabincell{c}{Comp.\\Ratio}}} \\
        \cmidrule(r){2-4}
        & linear & conv & embed & &  &  \\
        \midrule
        \multirow{5}{*}{Whisper} & $\times$ & $\times$ & $\times$ & 10.02 & 277.8 & - \\
        & $\checkmark$ & $\times$ & $\times$ & 10.46 & 130.8 & 2.12 \\
        & $\checkmark$ & $\checkmark$ & $\times$ & 10.60 & 127.8 & 2.17 \\
        & $\checkmark$ & $\times$ & $\checkmark$ & 11.37 & 42.2 & 6.58 \\
        & $\checkmark$ & $\checkmark$ & $\checkmark$ & 11.22 & 38.3 & 7.25 \\
        \midrule
        \multirow{5}{*}{Conformer} & $\times$ & $\times$ & $\times$ & 12.43 & 130.9 & - \\
        & $\checkmark$ & $\times$ & $\times$ & 12.51 & 31.6 & 4.14 \\
        & $\checkmark$ & $\checkmark$ & $\times$ & 12.69 & 23.4 & 5.59 \\
        & $\checkmark$ & $\times$ & $\checkmark$ & 12.61 & 27.3 & 4.79 \\
        & $\checkmark$ & $\checkmark$ & $\checkmark$ & 12.77 & 19.1 & 6.85 \\
        \bottomrule
    \end{tabular}
    \end{adjustbox}
    \label{tab:quantise}
    \vspace{-0.3cm}
\end{table}
First, WER and model compression ratios of systems after quantising different parts are shown in Table \ref{tab:quantise}. The majority of parameters in both the Whisper and Conformer models reside in the linear layers. Notably, applying NF4 quantisation to these layers has a very small impact on the performance. Furthermore, WER increased by 1.20\% for Whisper and only 0.34\% for Conformer upon NF4 quantisation. This suggests that models trained on smaller datasets are more robust to the quantisation noises under NF4 quantisation. In the following experiments, models with the highest compression ratios (i.e. the last row of each model in Table \ref{tab:quantise}) are used.  

\begin{table}[t]
\caption{\textit{WER on the LibriSpeech-SA and TL3-SA using quantised Whisper models. Whisper-baseline: Whisper after NF4 quantisation without adaptive training. FFT refers to full fine-tuning which trains all model parameters. Scratch refers to initialising LoRA weight randomly, and pretrain refers to the full PQM strategy with LoRA pretraining.}}
\vspace{0.2cm}
\begin{tabular*}{\columnwidth}{@{\extracolsep{\fill}}lcc}
\hline
\multicolumn{1}{c}{\multirow{2}{*}{\textbf{System}}} & \multicolumn{2}{c}{\textbf{WER(\%)}} \\
\multicolumn{1}{c}{} & LibriSpeech-SA & TL3-SA \\
\hline
Whisper-baseline & 11.22 & 7.71 \\
Whisper-FFT-FP32 & 9.59 & 6.85 \\
Whisper-FFT-NF4 & 10.51 & 7.19 \\
Whisper-LoRA-scratch-NF4 & 9.67 & 6.95 \\
Whisper-LoRA-pretrain-NF4 & \textbf{8.51} & \textbf{6.72} \\
\hline
\end{tabular*}
\label{tab:whisper}
\end{table}
Table \ref{tab:whisper} shows the performance of PQM on the Whisper base.en model. Compared to the baseline, the WER reduction achieved by fine-tuning all model parameters at full precision on target speaker data was largely reduced after model quantisation. As a result, the Whsper-FFT-NF4 model only achieved around 6.3\% relative WER reduction after speaker adaptive training. When LoRA was applied in conjunction with NF4 quantisation (i.e. the PQM strategy), without LoRA pretraining, the performance was already on par with the Whsper-FFT-FP32 full precision model and achieved a 13.8\% relative WER reduction on LibriSpeech-SA and a 9.9\% relative WER reduction on TL3-SA tasks. Moreover, when LoRA pretraining was applied in PQM, the improvements were further enlarged, with 24.2\% and 12.8\% relative WER reductions on LibriSpeech-SA and TL3-SA sets respectively. Note that the LoRA pretraining for TL3-SA was in fact cross-data, as the pretraining was done on the LibriSpeech clean-100 training set while directly applied to the TL3-SA data for adaptive training. This underscores the effectiveness of pretraining LoRA on data that resembles speaker-specific data.


\begin{table}[!ht]
\caption{\textit{WER on the LibriSpeech-SA using Conformer AED. Conformer-baseline: Conformer AED after NF4 quantisation without adaptive training. FFT, LoRA, scratch and pretrain follow the same definition as Table \ref{tab:whisper}}}
\vspace{0.2cm}
\centering
\begin{tabular}{lc}
\hline
\multicolumn{1}{c}{\textbf{System}} & \textbf{WER(\%)} \\
\hline
Conformer-baseline & 12.77 \\
Conformer-FFT-FP32 & 10.43 \\
Conformer-FFT-NF4 & 11.99 \\
Conformer-LoRA-scratch-NF4 & 10.25 \\
Conformer-LoRA-pretrain-NF4 & \textbf{9.54} \\
\hline
\end{tabular}
\label{tab:conformer}
\vspace{0.2cm}
\end{table}
The same set of experiments was also performed for the Conformer model as shown in Table \ref{tab:conformer}. Note that as the Conformer AED is trained on train-clean-100 already, we select 250 speakers from LibriSpeech train-clean-360 for LoRA pretraining. As before, the Conformer-LoRA-scratch-NF4 performed on par with the Conformer-FFT-FP32 full precision model, which achieved a 19.7\% relative WER reduction compared to the baseline. The Conformer-LoRA-pretrain-NF4 model achieved a further WER reduction, resulting in a relative 25.3\% WER reduction compared to the baseline.

\begin{figure}[t]
  \centering
  \includegraphics[width=1.08\linewidth]{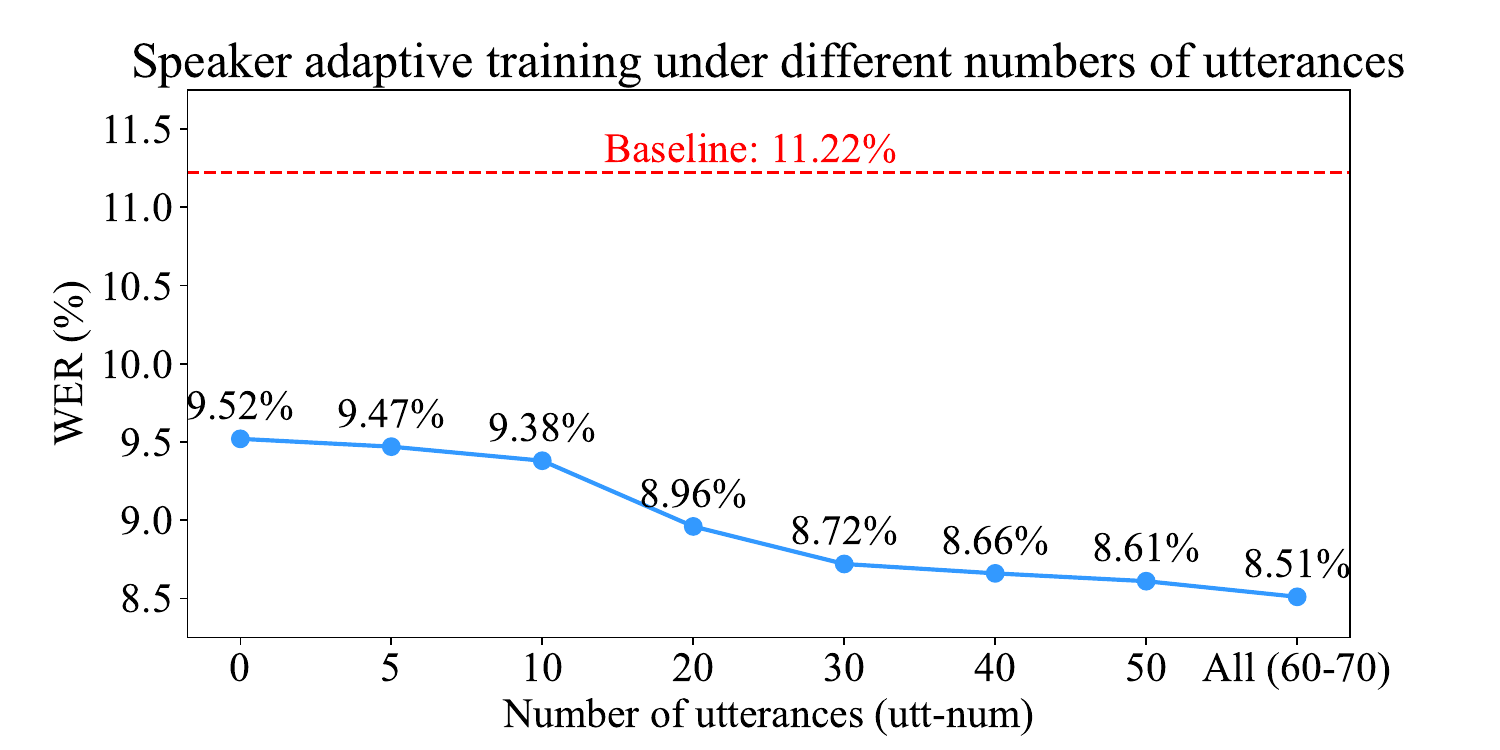}
  \caption{\textit{WER on the LibriSpeech-SA using Whisper-LoRA-pretrain-NF4 under different numbers of utterances. Utt-num=0 refers to Whisper-LoRA-pretrain-NF4 without speaker adaptive training.}}
  \label{fig:utt-num}
\end{figure}
Next, the influence of the number of training utterances was explored as shown in Fig. \ref{fig:utt-num}. It is evident that LoRA pretraining serves as a favourable starting point for speaker adaptive training. Moreover, the cost-effectiveness was maximized when 20 to 30 utterances were used for adaptive training with PQM.

\begin{table}[t]
\caption{\textit{WER on the LibriSpeech-SA using Whisper-LoRA-pretrain-NF4 and Conformer-LoRA-pretrain-NF4 under different label sources.}}
\vspace{0.2cm}
\centering
\resizebox{\columnwidth}{!}{%
\begin{tabular}{ccc}
\hline
\textbf{System} & \textbf{Label Source} & \textbf{WER(\%)} \\
\hline
\multirow{4}{*}{\parbox[t]{40.5mm}{\raggedright\arraybackslash Whisper-LoRA-pretrain-NF4}} 
 & No adaptive training & 9.52 \\
 & Ground truth & 8.51 \\
 & Whisper-large & 9.01 \\
 & Whisper-base.en & 9.44 \\
\hline
\multirow{4}{*}{\parbox[t]{40.5mm}{\raggedright\arraybackslash Conformer-LoRA-pretrain-NF4}}
 & No adaptive training & 11.81 \\
 & Ground truth & 9.54 \\
 & Whisper-large & 10.46 \\
 & Whisper-base.en & 10.71 \\
\hline
\end{tabular}
} 
\label{tab:semisup}
\vspace{-0.2cm}
\end{table}
Finally, to further alleviate the data scarcity issue, semi-supervised training for PQM was investigated for the Whisper and Conformer model and results are shown in Table \ref{tab:semisup}. As shown in the results, both Whisper-large and Whisper-base generated labels yielded improvements. Remarkably, with the training guided by Whisper-large labels showing, 5.4\% and 11.4\% relative WER reduction were achieved on the Whisper and Conformer model compared to without doing any speaker adaptive training. Note that this process did not require any human-annotated labels.


\section{Conclusions}
\label{sec:conclusions}
This paper proposes the PQM strategy to compensate for the performance loss due to model quantisation via personalisation. PQM adopted the NF4 quantisation approach together with LoRA-based speaker adaptive training and was applied to both the Conformer-based AED model and the Whisper model. Experiments on LibriSpeech and TL3 datasets using speaker adaptive training data partitions showed that personalisation largely improved the performance of quantised models. Specifically, using PQM, 15.1\% and 23.3\% relative WER reductions were achieved on quantised Whisper and Conformer-based AED models respectively, compared to the full precision models.

\vfill\pagebreak

\bibliographystyle{IEEEbib}
\bibliography{strings,refs}

\begin{thebibliography}{10}

\bibitem{radford2023robust}
A.~Radford, J.~W. Kim, T.~Xu, G.~Brockman, C.~McLeavey, and I.~Sutskever,
\newblock ``Robust speech recognition via large-scale weak supervision,''
\newblock in {\em Proc. ICML}, 2023.

\bibitem{gulati2020conformer}
A.~Gulati, J.~Qin, C.~Chiu, N.~Parmar, Y.~Zhang, J.~Yu, W.~Han, S.~Wang,
  Z.~Zhang, Y.~Wu, et~al.,
\newblock ``Conformer: {C}onvolution-augmented transformer for speech
  recognition,''
\newblock in {\em Proc. Interspeech}, 2020.

\bibitem{zhang2020transformer}
Q.~Zhang, H.~Lu, H.~Sak, A.~Tripathi, E.~McDermott, S.~Koo, and S.~Kumar,
\newblock ``Transformer transducer: {A} streamable speech recognition model
  with transformer encoders and {RNN-T} loss,''
\newblock in {\em Proc. ICASSP}, 2020.

\bibitem{kriman2020quartznet}
S.~Kriman, S.~Beliaev, B.~Ginsburg, J.~Huang, O.~Kuchaiev, V.~Lavrukhin,
  R.~Leary, J.~Li, and Y.~Zhang,
\newblock ``Quartznet: {D}eep automatic speech recognition with 1d time-channel
  separable convolutions,''
\newblock in {\em Proc. ICASSP}, 2020.

\bibitem{han2015deep}
S.~Han, H.~Mao, and W.~J. Dally,
\newblock ``Deep compression: {C}ompressing deep neural networks with pruning,
  trained quantization and {H}uffman coding,''
\newblock {\em arXiv preprint arXiv:1510.00149}, 2015.

\bibitem{qian2019binary}
Y.~Qian and X.~Xiang,
\newblock ``Binary neural networks for speech recognition,''
\newblock {\em Frontiers of Information Technology \& Electronic Engineering},
  vol. 20, no. 5, pp. 701--715, 2019.

\bibitem{leng2018extremely}
C.~Leng, Z.~Dou, H.~Li, S.~Zhu, and R.~Jin,
\newblock ``Extremely low bit neural network: Squeeze the last bit out with
  {ADMM},''
\newblock in {\em Proc. AAAI}, 2018.

\bibitem{nguyen2020quantization}
H.~D. Nguyen, A.~Alexandridis, and A.~Mouchtaris,
\newblock ``Quantization aware training with absolute-cosine regularization for
  automatic speech recognition,''
\newblock 2020.

\bibitem{ding20224}
S.~Ding, P.~Meadowlark, Y.~He, L.~Lew, S.~Agrawal, and O.~Rybakov,
\newblock ``4-bit conformer with native quantization aware training for speech
  recognition,''
\newblock {\em arXiv preprint arXiv:2203.15952}, 2022.

\bibitem{zhen2022sub}
K.~Zhen, H.~D. Nguyen, R.~Chinta, N.~Susanj, A.~Mouchtaris, T.~Afzal, and
  A.~Rastrow,
\newblock ``Sub-8-bit quantization aware training for 8-bit neural network
  accelerator with on-device speech recognition,''
\newblock {\em arXiv preprint arXiv:2206.15408}, 2022.

\bibitem{xu2022towards}
J.~Xu, S.~Hu, X.~Liu, and H.~Meng,
\newblock ``Towards green {ASR}: {L}ossless 4-bit quantization of a hybrid
  {TDNN} system on the 300-hr {S}witchboard corpus,''
\newblock {\em arXiv preprint arXiv:2206.11643}, 2022.

\bibitem{hernandez2023sharing}
S.~Hernandez, D.~Zhao, S.~Ding, A.~Bruguier, R.~Prabhavalkar, T.~N. Sainath,
  Y.~He, and I.~McGraw,
\newblock ``Sharing low rank conformer weights for tiny always-on ambient
  speech recognition models,''
\newblock in {\em Proc. ICASSP}, 2023.

\bibitem{yao2020int8}
Y.~Yao, Y.~Li, C.~Wang, T.~Yu, H.~Chen, X.~Jiang, J.~Yang, J.~Huang, W.~Lin,
  H.~Shu, et~al.,
\newblock ``Int8 {Winograd} acceleration for conv1d equipped asr models
  deployed on mobile devices,''
\newblock {\em arXiv preprint arXiv:2010.14841}, 2020.

\bibitem{dettmers2023qlora}
T.~Dettmers, A.~Pagnoni, A.~Holtzman, and L.~Zettlemoyer,
\newblock ``{QLoRA}: {E}fficient finetuning of quantized {LLMs},''
\newblock {\em arXiv preprint arXiv:2305.14314}, 2023.

\bibitem{hu2021lora}
E.~J. Hu, Y.~Shen, P.~Wallis, Z.~Allen-Zhu, Y.~Li, S.~Wang, L.~Wang, and
  W.~Chen,
\newblock ``{LoRA}: {L}ow-rank adaptation of large language models,''
\newblock {\em arXiv preprint arXiv:2106.09685}, 2021.

\bibitem{tensorflow_lite}
{G. Inc.},
\newblock ``{{TensorFlow Lite}: {A}n open source deep learning framework for
  on-device inference},'' 2017,
\newblock Accessed: 2023-09-01.

\bibitem{jiang2020mnn}
X.~Jiang, H.~Wang, Y.~Chen, Z.~Wu, L.~Wang, B.~Zou, Y.~Yang, Z.~Cui, Y.~Cai,
  T.~Yu, et~al.,
\newblock ``{MNN}: {A} universal and efficient inference engine,''
\newblock {\em Proceedings of Machine Learning and Systems}, vol. 2, pp. 1--13,
  2020.

\bibitem{ncnn}
{T. Inc.},
\newblock ``{{NCNN}: {H}igh-performance neural network inference computing
  framework optimized for mobile platforms},'' 2017,
\newblock Accessed: 2023-09-01.

\bibitem{dettmers20218}
T.~Dettmers, M.~Lewis, S.~Shleifer, and L.~Zettlemoyer,
\newblock ``8-bit optimizers via block-wise quantization,''
\newblock {\em arXiv preprint arXiv:2110.02861}, 2021.

\bibitem{saon2013speaker}
G.~Saon, H.~Soltau, D.~Nahamoo, and M.~Picheny,
\newblock ``Speaker adaptation of neural network acoustic models using
  i-vectors,''
\newblock in {\em Proc. ASRU}, 2013.

\bibitem{sari2020unsupervised}
L.~Sar{\i}, N.~Moritz, T.~Hori, and J.~Le~Roux,
\newblock ``Unsupervised speaker adaptation using attention-based speaker
  memory for end-to-end {ASR},''
\newblock in {\em Proc. ICASSP}, 2020.

\bibitem{rohdin2018end}
J.~Rohdin, A.~Silnova, M.~Diez, O.~Plchot, P.~Mat{\v{e}}jka, and L.~Burget,
\newblock ``End-to-end {DNN} based speaker recognition inspired by i-vector and
  {PLDA},''
\newblock in {\em Proc. ICASSP}, 2018.

\bibitem{yue2020autoencoder}
Z.~Yue, H.~Christensen, and J.~Barker,
\newblock ``Autoencoder bottleneck features with multi-task optimisation for
  improved continuous dysarthric speech recognition,''
\newblock in {\em Proc. Interspeech}, 2020.

\bibitem{cardinal2015speaker}
P.~Cardinal, N.~Dehak, Y.~Zhang, and J.~Glass,
\newblock ``Speaker adaptation using the i-vector technique for bottleneck
  features,''
\newblock in {\em Proc. Interspeech}, 2015.

\bibitem{doddipatla2014speaker}
R.~Doddipatla, M.~Hasan, and T.~Hain,
\newblock ``Speaker dependent bottleneck layer training for speaker adaptation
  in automatic speech recognition,''
\newblock in {\em Proc. Interspeech}, 2014.

\bibitem{swietojanski2016learning}
P.~Swietojanski, J.~Li, and S.~Renals,
\newblock ``Learning hidden unit contributions for unsupervised acoustic model
  adaptation,''
\newblock {\em IEEE/ACM Transactions on Audio, Speech, and Language
  Processing}, vol. 24, no. 8, pp. 1450--1463, 2016.

\bibitem{zhang2016dnn}
C.~Zhang and P.~C. Woodland,
\newblock ``{DNN} speaker adaptation using parameterised sigmoid and {ReLU}
  hidden activation functions,''
\newblock in {\em Proc. ICASSP}, 2016.

\bibitem{miao2015speaker}
Y.~Miao, H.~Zhang, and F.~Metze,
\newblock ``Speaker adaptive training of deep neural network acoustic models
  using i-vectors,''
\newblock {\em IEEE/ACM Transactions on Audio, Speech, and Language
  Processing}, vol. 23, no. 11, pp. 1938--1949, 2015.

\bibitem{swietojanski2016sat}
P.~Swietojanski and S.~Renais,
\newblock ``{SAT-LHUC}: {S}peaker adaptive training for learning hidden unit
  contributions,''
\newblock in {\em Proc. ICASSP}, 2016.

\bibitem{liao2013speaker}
H.~Liao,
\newblock ``Speaker adaptation of context dependent deep neural networks,''
\newblock in {\em Proc. ICASSP}, 2013.

\bibitem{yu2013kl}
D.~Yu, K.~Yao, H.~Su, G.~Li, and F.~Seide,
\newblock ``{KL}-divergence regularized deep neural network adaptation for
  improved large vocabulary speech recognition,''
\newblock in {\em Proc. ICASSP}, 2013.

\bibitem{whisperbiasing}
{G. Sun, X. Zheng, C. Zhang, P. C. Woodland},
\newblock ``Can contextual biasing remain effective with {Whisper} and
  {GPT}-2?,''
\newblock in {\em Proc. Interspeech}, 2023.

\bibitem{espnet}
S.~Watanabe, T.~Hori, S.~Karita, T.~Hayashi, J.~Nishitoba, Y.~Unno, N.~{E. Y.
  Soplin}, J.~Heymann, M.~Wiesner, N.~Chen, A.~Renduchintala, and T.~Ochiai,
\newblock ``{ESPnet}: {E}nd-to-end speech processing toolkit,''
\newblock in {\em Proc. Interspeech}, 2018.

\end{thebibliography}

\end{document}